\documentclass{article}
\usepackage{graphicx}

\begin{document}

%New Commands
\newcommand{\siml}{\stackrel{<}{\sim}}
\newcommand{\simg}{\stackrel{>}{\sim}}
\newcommand{\lleq}{\stackrel{<}{=}}

\baselineskip=1.333\baselineskip
%PRE
%\baselineskip=2.0\baselineskip

%\draft
%\baselineskip=0.5\baselineskip

%
\begin{center}
{\large\bf
A Generalized Rate Model
for Neuronal Ensembles
\footnote{in {\it Neuronal Network Research Horizons},
Ed. M. L. Weiss, Nova Science Publishers, 2007}
} 
\end{center}

\begin{center}
Hideo Hasegawa
\footnote{Electronic address:  hasegawa@u-gakugei.ac.jp}
\end{center}

\begin{center}
{\it Department of Physics, Tokyo Gakugei University  \\
Koganei, Tokyo 184-8501, Japan}
\end{center}
\begin{center}
%{\rm (Jan. 11, 2005)}
({\today})
\end{center}
%\maketitle
\thispagestyle{myheadings}

\begin{abstract}
There has been a long-standing controversy whether information 
in neuronal networks is carried by the firing rate code or 
by the firing temporal code. The current status of the rivalry 
between the two codes is briefly reviewed with the recent studies
such as the brain-machine interface (BMI). 
Then we have proposed a generalized rate model
based on the {\it finite}
$N$-unit Langevin model subjected to
additive and/or multiplicative noises, 
in order to understand the firing property 
of a cluster containing $N$ neurons.
The stationary property of
the rate model has been studied with the use of the Fokker-Planck 
equation (FPE) method. Our rate model is shown to yield various 
kinds of stationary distributions such as the interspike-interval 
distribution expressed by non-Gaussians including gamma, 
inverse-Gaussian-like and log-normal-like distributions.

The dynamical property of the generalized
rate model has been studied with the use
of the augmented moment method (AMM) which was
developed by the author 
[H. Hasegawa, J. Phys. Soc. Jpn. 75 (2006) 033001].
From the macroscopic point of view in the AMM, 
the property of the $N$-unit neuron cluster is expressed
in terms of {\it three} quantities; $\mu$, the mean of spiking rates of
$R=(1/N) \sum_i r_i$ where $r_i$ denotes the firing rate 
of a neuron $i$ in the cluster:
$\gamma$, averaged fluctuations in local variables ($r_i$):
$\rho$, fluctuations in global variable ($R$).
We get equations of motions of the three quantities, which show
$\rho \sim \gamma/N$ for weak couplings. This implies
that the population rate code is generally more reliable than
the single-neuron rate code. 
Dynamical responses of the neuron cluster to pulse and sinusoidal inputs
calculated by the AMM are in good agreement
with those by direct simulations (DSs).

Our rate model is extended and applied to
an ensemble containing multiple neuron clusters.
In particular, we have studied the property of
a generalized Wilson-Cowan model for
an ensemble consisting of
two kinds of excitatory and inhibitory clusters.

%The relation between the result of AMM 
%and the maximum a posteriori
%(MAP) estimate based on the Bayesian statistics
%is discussed also.

%An application of the AMM 
%to the FitzHugh-Nagumo model,
%a typical spiking neuron model, subjected 
%to multiplicative noises is also presented
%with some numerical results.

\end{abstract}

%\noindent
%\vspace{0.5cm}

%{\it PACS No.} 05.10.Gg, 05.45.-a, 84.35.+i
%
\vspace{1cm}

%\vspace{2cm}

%{\Large\bf EDITOR: Please review the English.}

\newpage

\noindent
{\bf Contents}

\noindent
1. Introduction ……………………………………………………………………………… 3

\vspace{0.5cm}

\noindent
2. Single neuron clusters ......................................................................................... 8

2.1 Rate model ................................................................................................... 8

2.2 Stationary property ...................................................................................... 9

2.3 Dynamical property ......................................................................................13

2.3.1 AMM .......................................................................................................13

2.3.2 Response to pulse inputs ..................................................................................14

2.3.3 Response to sinusoidal inputs ..........................................................................15

\vspace{0.5cm}

\noindent
3. Multiple neuron clusters ........................................................................................16

3.1 AMM ......................................................................................................................16

3.2 Stationary property ...............................................................................................16

3.3 Dynamical property ...............................................................................................19

\vspace{0.5cm}

\noindent
4. Discussion and conclusion .......................................................................................20
\vspace{0.5cm}

\noindent
Acknowledgements
\vspace{0.5cm}

\noindent
Appendix  AMM for multiple clusters .......................................................................23

\vspace{0.5cm}

\noindent
References .....................................................................................................................25

\vspace{1.0cm}
\noindent
{\bf Abbreviations}

AMM: augmented moment method

BCI: brain-computer interface

BMI: brain-machine interface

DS: direct simulation

FPE: Fokker-Planck equation

\newpage
%\narrowtext
\section{Introduction}

%\subsection{Rate code vs. temporal code}

Human brain contains more than $10^{10}$ neurons \cite{Kandel95}.
Neurons communicate information, emitting 
short voltage pulses called spikes, 
which propagate through axons and dendrites
to neurons at the next stage.
It has been a long-standing controversy how neurons communicate 
information by firings or spikes \cite{Rieke96}-\cite{deCharms00}.
The issue on the neural code is whether information is encoded in the
rate of firings of neurons ({\it rate code}) or in
the more precise firing times ({\it temporal code}).
The rate code was first recognized by Adrian \cite{Adrian26} who 
noted that the neural firing was increased 
with increasing the stimulus intensity. 
The firing rate $r(t)$ of a neuron is defined by
\begin{eqnarray}
r(t)&=&\frac{n(t, t+t_w)}{t_w}
= \frac{1}{t_w} \int_{t}^{t+t_w} \sum_k
\delta(t' - t_{k})\: dt',
\end{eqnarray}
where $t_w$ denotes the time window,
$n(t, t+t_w)$ the number of firings between
$t$ and $t+t_w$, and $t_{k}$ the {\it k}th firing time. 
It has been widely reported that
firing activities of motor and sensory neurons
vary in response to applied stimuli.
In the temporal code, on the contrary,
the temporal information like
the inter-spike interval (ISI) defined by
\begin{eqnarray}
T_{k}&=& t_{k+1}-t_{k},
\end{eqnarray}
and its distribution $\pi(T)$, ISI histogram (ISIH),
are considered to play
the important role in the neural activity.
More sophisticate methods such 
as the auto- and cross-correlograms
and joint peri-stimulus time histogram (JPSTH)
have been also employed for an analysis of
correlated firings based on the temporal-code hypothesis.
There have been accumulated
experimental evidences, which seem to indicate a use of
the temporal code in brain \cite{Rieke96}\cite{Ikegaya04}-\cite{Shmiel05}.
A fly can react to new stimuli and the change the direction of flight
within 30-40 ms \cite{Rieke96}.
In primary visual cortices of mouse and cat,
repetitions of spikes with the millisecond accuracy have been
observed \cite{Ikegaya04}.
Humans can recognize visual scenes within tens of milliseconds,
even though recognition is involved several processing steps 
\cite{Thorpe96}-\cite{Shmiel05}.

%\subsection{Single-neuron code vs. population code}

The other issue on the neural code is whether information
is encoded in the activity of
a single (or very few) neuron or 
that of large number of neurons 
({\it population code})\cite{Abbott98}\cite{Pouget00}.
The classical population code is employed
in a number of systems
such as motor neurons \cite{Georg86}, 
the place cell in hippocampus \cite{Hippo} 
and neurons in middle temporal (MT) areas \cite{Treue00}.
Neurons in motor cortex of the monkey, for example, encode
the direction of reaching movement of the arm.
Information on the direction of the arm movement
is decoded from a set of neuronal firing rates by
summing the preferred direction vector weighted by
the firing rate of each neuron in the neural population \cite{Georg86}. 
%Recently, a probabilistic population code based on 
%the Bayesian statistics has been proposed \cite{Zemel98}. 
It has been considered
that the rate code needs the time window of a few hundred milliseconds
to accumulate information on firing rate, 
and then its information
capacity and speed are limited compared to the temporal code.
It is, however, not true when
the rate is averaged over the ensembles
({\it population rate code}).
The population average promotes the response, suppressing
the effects of neuronal variability
\cite{Shadlen94}\cite{Abbott99}. 
The population rate code is expected to be a useful coding method
in many areas in the brain.
Indeed, in recent brain-machine interface (BMI) 
or brain-computer interface (BCI)
\cite{Chapin99}\cite{Carmena03}\cite{Anderson04},
the information on hand position and velocity {\it etc.}
at time $t$ is modeled as a weighted linear combination of neuronal
firing rates collected by multi-electrodes 
as given by \cite{Carmena03}
\begin{equation}
{\bf y}(t)= \sum_u {\bf a}(u) \: {\bf r}(t-u).
\end{equation}
Here ${\bf y}(t)$ denotes a vector expressing
position, velocity {\it etc.}, ${\bf r}(t-u)$ expresses
a vector of firing rates at time $t$ and timelag $u$, and
${\bf a}(u)$ stands for a vector of weight at the timelag $u$. 
% and ${\bf b}$ means a vector of $y$-intercepts.
Equation (3) is transformed to a matrix form, from which
a vector ${\bf a(u)}$ is obtained
by the linear filter method \cite{Warland97}\cite{Stanley99}
with training data of ${\bf y}(t)$ and ${\bf r}(t-u)$.
The predictor of a new, unobserved stimulus
${\bf \hat{y}}(t)$ for a observed 
${\bf \hat{r}}(t-u)$ is then given by
\begin{equation}
{\bf \hat{y}}(t)= \sum_u {\bf a}(u) \: {\bf \hat{r}}(t-u).
\end{equation}
Visual and sensory feedback signals like the pressure on
animal's skin may be sent back to the brain.
It has been reported that an artificial hand
is successfully manipulated by such decoding method
for observed firing-rate signals of ${\bf \hat{r}(t-u)}$
\cite{Chapin99}\cite{Carmena03}.
A success in BMI strongly suggests that the population
rate code is employed in
sensory and motor neurons while it is still not clear
which code is adopted in higher-level cortical neurons.

%\subsection{Spiking model}

The microscopic, conductance-based mechanism of 
firings of neurons is fairly well understood.
The dynamics of the membrane potential $V_i$ 
of the neuron $i$ in the neuron cluster
is expressed by the Hodgkin-Huxley-type model given by  \cite{Gerstner02}
\begin{eqnarray}
C \frac{d V_i}{d t} &=& \sum_n g_n(V_i, \alpha_n)
(V_r-V_i) +I_i.
\end{eqnarray}
Here $C$ expresses the capacitance of a cell:
$V_r$ is the recovery voltage:
$g_n(V_i, \alpha_n)$ denotes the conductance for $n$ ion channel
($n=$Na, K {\it etc.}) which
depends on $V_i$ and $\alpha_n$, the gate function 
for the channel $n$: $I_i$ is
the input arising from couplings with other neurons and
an external input.
The dynamics of $\alpha_n$ is expressed by the first-order
differential equation.
Since it is difficult to solve nonlinear differential equations
given by Eq. (5), the reduced, simplified neuron models
such as the integrate-and-fire (IF),
the FitzHugh-Nagumo (FN)
and Hindmarsh-Rose (HR) models have been employed.
In the simplest IF model
with a constant conductance,
Eq. (5) reduces to the linear differential 
equation, which requires an artificial reset of $V_i$ at 
the threshold $\theta$.

It is not easy to analytically obtain the rate $r(t)$
or ISI $T(t)$ from spiking neuron model given by Eq. (5).
There are two approaches in extracting
the rate from spiking neuron model.
In the first approach, Fokker-Planck equation (FPE)
is applied to the IF model
to calculate the probability $p(V, t)$,
from which the rate $r(t)$ is obtainable.
In order to avoid the difficulty of the
range of variable:
$-\infty < V \leq \theta$ in the original IF model,
sophisticate quadratic and exponential
IF models, in which the range of variable is
$-\infty < V \leq \infty$,
have been proposed \cite{Trocme05}.
In the second approach, the rate model
is derived from the conductance-based model
with synapses by using
the $f-I$ relation between
the applied dc current $I$ and the frequency
$f$ of autonomous firings
\cite{Amit91}\cite{Ermen94}\cite{Shriki03}.
It has been shown that the conductance-based
model may lead to the rate model
if the network state does not posses a high degree of
synchrony \cite{Shriki03}.

%\subsection{Rate model}

In the rate-code hypothesis, 
a neuron is regarded as a black box
receiving and emitting signals expressed by the firing rate. 
The dynamics of
the rate $r_i(t)$ of the neuron $i$ in a cluster
is expressed by 
\begin{eqnarray}
\tau \frac{d r_i}{d t} &=& - r_i 
+ K\left(\sum_j w_{ij} r_j + I_i \right),
\end{eqnarray}
where $\tau$ denotes the relaxation time,
$w_{ij}$ the coupling between neurons $i$ and $j$,
and $I_i$ an external input.
The gain function $K(x)$ is usually given by
the sigmoid function or by the $f-I$ relation
mentioned above.
One of disadvantages of the rate model is
that the mechanism is not well biologically supported.
Nevertheless, the rate model has been adopted
for a study on many subjects of the brain.
The typical rate model is the Wilson-Cowan (WC) model,
with which the stability of a cluster consisting
of excitatory and inhibitory neurons is 
investigated \cite{Wilson72}\cite{Amari72}.
The rate model given by Eq. (6) with $K(x)=x$
is the Hopfield model \cite{Hopfield84},
which has been extensively employed 
for a study on the memory in the brain
with incorporating the plasticity of synapses into $w_{ij}$.

%\subsection{Noises in brains}

It is well known that neurons in brains are subjected to
various kinds of noises, though their precise origins
are not well understood. 
The response of neurons to stimuli is modified by noises in various ways.
Although firings of a single {\it in vitro} neuron 
are reported to be
precise and reliable \cite{Mainen95}, 
those of {\it in vivo} neurons are quite
unreliable, which is expected to be due to noisy environment.
The strong criticism against the temporal code is that 
it is not vulnerable to noises, while the rate code
is robust against them.

Noises may be, however, beneficial for the signal 
transmission in the brain against our wisdom.
The most famous phenomenon is
the stochastic resonance \cite{Gamma98}, 
in which the signal-to-noise
ratio of subthreshold signals is improved by noises. 
It has been shown that the noise is essential for the rate-code
signal to propagate through multilayers described by the IF model:
otherwise firings tend to synchronize, by which the rate-code signal
is deteriorated \cite{Masuda02}\cite{Rossum02}. 
Recent study using HH model has shown that
firing-rate signals propagate through the multiplayer
with the synchrony \cite{Wang06}.

It is theoretically supposed that there are two types of noises:
additive and multiplicative noises.
The magnitude of the former is independent of the
state of variable while that of the latter
depends on its state.
Interesting phenomena caused by the two noises have
been investigated \cite{Munoz04}.
It has been realized that
the property of multiplicative noises
is different from that of additive noises
in some respects.
(1) Multiplicative noises induce the phase transition,
creating an ordered state, while additive noises
are against the ordering \cite{Munoz05}\cite{Broeck94}.
%(2) Although the stochastic resonance is not realized
%in linear systems with additive noises,
%it may be possible with multiplicative {\it color} noise
%(but not with multiplicative {\it white} noise) 
%\cite{Berd96,Bazykin97}.
(2) Although the probability distribution in stochastic systems
subjected to additive Gaussian noise follows the Gaussian,
it is not the case for multiplicative
Gaussian noises which generally yield non-Gaussian distribution
\cite{Tsallis88}-\cite{Hasegawa05b}.
(3) The scaling relation of the effective 
strength for additive noise given by
$\beta(N)=\beta(1)/\sqrt{N}$ is not applicable to
that for multiplicative noise:
$\alpha(N) \neq \alpha(1)/\sqrt{N}$, where $\alpha(N)$ and $\beta(N)$
denote effective strengths of multiplicative
and additive noises, respectively, in the $N$-unit system
\cite{Hasegawa06a}.
A naive approximation
of the scaling relation for multiplicative noise: 
$\alpha(N)=\alpha(1)/\sqrt{N}$
as adopted in Ref. \cite{Munoz05}, yields the result
which does not agree with that of direct simulation (DS) 
\cite{Hasegawa06a}.

%\subsection{Stochastic model}

Formally, noise can be introduced to the 
spiking and rate models, by adding
a fluctuating noise term on the right-hand side
of Eqs. (5)  and (6), respectively.
ISI data obtained from neuronal
systems have been analyzed by using
various population rate methods (for a recent review, 
see Refs. \cite{Kass05,Vog05}).
Experimental ISI data cannot always be described in terms of
a single probability distribution.
They are fitted 
by a superposition of
some known probability densities such as
the gamma, inverse-Gaussian and log-normal distributions.
The gamma distribution 
with parameters $\lambda$ and $\mu$
is given by
\begin{equation}
P_{gam}(x)=\frac{\mu^{-\lambda}}{\Gamma(\lambda)}\:x^{\lambda-1}
\exp\left(-{\frac{x}{\mu}} \right),
\end{equation}
which is derived from a simple 
stochastic integrate-and-fire (IF) model 
with additive noises
for Poisson inputs \cite{Tuckwell88},
$\Gamma(x)$ being the gamma function.
For $\lambda=1$ in Eq. (7), we get
the exponential distribution
relevant to a simple Poisson process.
The inverse Gaussian distribution 
with parameters $\lambda$ and $\mu$ given by
\begin{equation}
P_{IG}(x)=\left( \frac{\lambda}{2 \pi x^3} \right)^{1/2}
\exp \left[ -\frac{\lambda (x-\mu)^2}{2 \mu^2 x} \right],
\end{equation}
is obtained
from a stochastic IF model in which
the membrane potential is represented as a random walk
with drift \cite{Gerstein64}.  
The log-normal distribution 
with parameters $\mu$ and $\sigma$ given by
\begin{equation}
P_{LN}(x)=\frac{1}{\sqrt{2 \pi \sigma^2 }\:x}
\exp \left[- \frac{(\log x -\mu)^2}{2 \sigma^2}  \right],
\end{equation}
is adopted when
the log of ISI is assumed to follow the Gaussian
\cite{McKeegan02}.
Fittings of experimental ISI data to a superposition
of these probability densities have been extensively
discussed in the literature \cite{Kass05}-\cite{McKeegan02}.

Much study has been made on the spiking neuron model 
for coupled ensembles
with the use of two approaches: direct simulations (DSs)
and analytical approaches such as FPE and moment method.
DSs have been performed for large-scale
networks mostly consisting of IF neurons.
Since the time to simulate networks by conventional methods 
grows as $N^2$ 
with $N$, the size of the network,
it is rather difficult to simulate realistic neuron clusters.
In the FPE, dynamics of neuron ensembles
is described by the population activity.
Although the FPE is powerful method which is formally
applicable to the arbitrary $N$, actual calculations have been
made for $N=\infty$ with the
mean-field and diffusion approximations.  
Quite recently, the population density method has been developed
as a tool modeling large-scale neuronal clusters 
\cite{Omurtag00},\cite{Haskell00}. 
As a useful semi-analytical method for stochastic neural models,
the moment method was proposed\cite{Rod96}. 
For example,
when the moment method is applied to $N$-unit FN model,
original $2N$-dimensional
stochastic equations are transformed to $N(2 N+3)$-dimensional
deterministic equations.
This figure becomes 230, 20300 and 2 003 000
for $N=10$, 100 and 100, respectively.

%\subsection{Augmented moment method}

In many areas of the brain, neurons are organized into 
groups of cells such as column in the visual cortex \cite{Mount57}.
Experimental findings have shown that
within small clusters consisting of finite number of
neurons ($\sim 10-1000$), there exist
many cells with very nearly identical responses 
to identical stimuli \cite{Mount57}.
Analytical, statistical methods 
having been developed so far %for an analysis of neurons ensembles
are mean-field-type theories which may deal with
infinite-size systems, but not with finite-size ones.
Based on a macroscopic point of view,
Hasegawa \cite{Hasegawa03a} has
proposed the augmented moment method (AMM),
which emphasizes not the property of individual neurons 
but rather that of ensemble neurons.
In the AMM,
the state of {\it finite} $N$-unit stochastic ensembles 
is described by a fairly small number of variables:
averages and fluctuations of local and global variables.
For $N$-unit FN neuron ensembles, for example,
the number of deterministic equation in the AMM 
becomes {\it eight}
independent of $N$. This figure of eight is much smaller that
those in the moment method mentioned above.
The AMM has been successfully
applied to a study on the dynamics of 
the Langevin model and stochastic spiking models
such as FN and HH models, with global, local 
or small-world couplings
(with and without transmission delays)
\cite{Hasegawa03b}-\cite{Hasegawa05a}. 

The AMM in Ref. \cite{Hasegawa03a} was originally
developed by expanding variables
around their stable mean values in order to obtain
the second-order moments both for
local and global variables in stochastic systems.  
In recent papers \cite{Hasegawa06a}\cite{Hasegawa06b},  
we have reformulated the AMM with the use of FPE
to discuss stochastic systems subjected to multiplicative noise:
the FPE is adopted to avoid the difficulty
due to the Ito versus Stratonovich calculus inherent
to multiplicative noise \cite{Risken96}. 

An example of calculations with the use of the AMM for
a FN neuron cluster subjected to additive and
multiplicative noises is presented in Figs. 1(a)-(e)
\cite{Hasegawa06c}.
When input pulses shown in Fig. 1(a) are applied to
the 10-unit FN neuron cluster, 
the membrane potential $v_i(t)$ of a given neuron
depicted in Fig. 1(b) is obtained by DS with a single trial.
It has the much irregularity because of added noises.
%with $\alpha=0.1$ and $\beta=0.2$ ($w=0.5$).
When we get the ensemble-averaged potential given by
$V(t)=(1/N)\sum_i v_i(t)$, 
the irregularity
is reduced as shown in Fig. 1(c).  This is
one of the advantages of the population code 
\cite{Shadlen94}.  
The results shown in Figs. 1(b) and 1(c)
are obtained by a single trial.
When we have repeated DS and taken the average over 100 trials,
the irregularity in $V(t)$ is furthermore reduced as shown in Fig. 1(d). 
The result of the AMM, $\mu(t)$, plotted in Fig. 1(e) is
in good agreement with that of DS in Fig. 1(d).

The purpose of the present paper is two fold:
to propose a generalized rate model for 
neuron ensembles and to study its property
with the use of FPE and the AMM.
The paper is organized as follows.
In Sec. 2, we discuss the generalized rate model for 
a single cluster containing $N$ neurons,
investigating its stationary and
dynamical properties.
In Sec. 3, our rate model is extended and applied to an ensemble
containing multiple neuron clusters. In particular,
we study the two-cluster ensemble consisting of
excitatory and inhibitory clusters.
The final Sec. 4 is devoted to discussion and
conclusion. 

%\newpage
\section{Single neuron clusters}

\subsection{Generalized rate model}

We have adopted a neuronal cluster consisting of
$N$ neurons.
The dynamics of firing rate $r_{i}$ ($\geq 0$) of a neuron $i$
is assumed to be described by the Langevin model given by
\begin{eqnarray}
\frac{dr_{i}}{dt} &=& F(r_{i}) 
+H(u_{i})
+ \alpha G(r_{i}) \eta_{i}(t)+ \beta \xi_{i}(t),
\hspace{1cm}\mbox{($i=1-N$)}
\end{eqnarray}
with 
\begin{eqnarray}
u_{i}(t) &=& \left( \frac{w}{Z} \right) 
\sum_{k (\neq i)} r_{k}(t)
+ I_i(t), 
\end{eqnarray}
where $F(x)$, $G(x)$ and $H(x)$ are arbitrary functions of $x$:
%$H(x)$ the sigmoid function with the threshold $\theta_m$ and
%width $d_m$ for the ensemble $m$: 
$Z$ $(=N-1)$ denotes the coordination number: 
$w$ is the coupling strength: 
$I_i(t)$ expresses an input from external sources:
$\alpha$ and $\beta$ are the strengths of additive and
multiplicative noises, respectively, given by 
$\eta_i(t)$ and $\xi_i(t)$ expressing zero-mean Gaussian white
noises with correlations given by
\begin{eqnarray}
<\eta_{i}(t)\:\eta_{j}(t')> 
&=& \delta_{ij} \delta(t-t'),\\
<\xi_{i}(t)\:\xi_{j}(t')> 
&=& \delta_{ij} \delta(t-t'),\\
<\eta_{i}(t)\:\xi_{j}(t')> &=& 0.
\end{eqnarray}

The rate models proposed so far have employed
$F(x)=-\lambda x$ and $G(x)=0$ (no multiplicative noises).
In this paper, we will adopt several functional forms
for $F(x)$ and $G(x)$.
As for the gain function $H(x)$, 
two types of expressions have been adopted.
In the first category, the sigmoid function such as
$H(x)={\rm tanh}(x)$,
$1/(1+{\rm e}^{-x})$,
${\rm atan}(x)$, {\it etc.} have been adopted.
In the second category, $H(x)$ is given by
the $f-I$ function
as $H(x)=(x-x_c) \Theta(x-x_c)$ which
expresses the frequency $f$ of autonomous oscillation
of a neuron
for the applied dc current $I$,
$x_c$ denoting the critical value
and $\Theta(x)$ the Heaviside function:
$\Theta(x)=1$ for $x \geq 0$ and 0 otherwise.
It has been theoretically shown in Ref. \cite{Hasegawa00a} 
that when spike inputs with the mean ISI 
($T_i$) are applied to
an HH neuron, the mean ISI of output signals ($T_o$) 
is $T_o = T_i$ for $T_i \siml 15 $ ms
and $T_o \sim 15$ for $T_i > 15 $ ms.
This is consistent with the recent 
calculation for HH neuron
multilayers, which shows a nearly linear
relationship between the input ($r_i$) 
and output rates ($r_o$) 
for $r_i < 60$ Hz \cite{Wang06}.
It is interesting that
the $r_i-r_o$ relation is continuous
despite the fact that the HH neuron
has the discontinuous first-type $f-I$ relation.
We will adopt, in this paper, 
a simple expression given by
\cite{Monteiro02}
\begin{eqnarray}
H(x) &=& \frac{x}{\sqrt{x^2+1}}, 
\end{eqnarray}
although our result is valid
for an arbitrary form for $H(x)$.
The nonlinear, saturating behavior in $H(x)$ 
arises from the property of the refractory period ($\tau_r$)
where spike outputs are prevented for 
$t_f < t < t_f+\tau_r$ after firing at $t=t_f$.

\subsection{Stationary property}
\subsubsection{Distribution of $r$}

The Fokker-Planck equation for the distribution of
$p(\{r_{i} \},t)$
is given by 
\cite{Haken83}
\begin{eqnarray}
\frac{\partial}{\partial t}\: p(\{r_{i} \},t)&=&
-\sum_{k} \frac{\partial}{\partial r_{k}}\{ [F(r_{k}) +
\frac{\phi\alpha^2}{2}G'(r_{k})G(r_{k}) 
+ H(u_{k})]\:p(\{ r_{i} \},t)\}  
\nonumber \\
&&+\frac{1}{2}\sum_{k}\frac{\partial^2}{\partial r_{k}^2} 
\{[\alpha^2 G(r_{k})^2+\beta^2]\:p(\{ r_{i} \},t) \},
\end{eqnarray}
where $G'(x)=dG(x)/dx$, and
$\phi=1$ and 0 in the Stratonovich and Ito representations,
respectively.

The stationary distribution $p(r)$ for $w=0$
and $I_i(t)=I$ 
is given by
\begin{eqnarray}
{\rm ln}\:p(r) &\propto& X(r)+Y(r)
-\left( 1-\frac{\phi}{2} \right){\rm ln} 
\left[\frac{\alpha^2 G(r)^2}{2}+\frac{\beta^2}{2} \right],
\end{eqnarray}
with
\begin{eqnarray}
X(r) &=& 2 \int \:dr \:
\left[ \frac{F(r)}{\alpha^2 G(r)^2+\beta^2} \right], \\
Y(r) &=& 2 \int \:dr \:
\left[ \frac{H(I)}{\alpha^2 G(r)^2+\beta^2} \right].
\end{eqnarray}
Hereafter we mainly adopt the Stratonovich representation.

\noindent
{\bf Case I} $F(x)=-\lambda x$ and $G(x)=x$

For the linear Langevin model, we get
\begin{eqnarray}
p(r) &\propto& 
\left[1+\left( \frac{\alpha^2 r^2}{\beta^2} \right)
\right]^{-(\lambda/\alpha^2+1/2)}
\:e^{Y(r)}, 
\end{eqnarray}
with
\begin{eqnarray}
Y(r)=\left( \frac{2 H}{\alpha \beta} \right)
{\rm arctan}\left( \frac{\alpha r}{\beta} \right),
\end{eqnarray}
where $H=H(I)$.
In the case of $H=Y(r)=0$, we get
the $q$-Gaussian given by
\cite{Sakaguchi01,Anten02}
\begin{eqnarray}
p(r) &\propto& \left[ 1-(1-q)\gamma r^2 \right]^{\frac{1}{1-q}},
\end{eqnarray}
with
\begin{eqnarray}
\gamma&=& \frac{2 \lambda+\alpha^2}{2 \beta^2}, \\
q&=& \frac{2 \lambda+3 \alpha^2}{2 \lambda +\alpha^2}.
\end{eqnarray}
We examine the some limiting cases of Eq. (20) as follows.

\noindent
(A) For $\alpha=0$ and $\beta \neq 0$, Eq. (20) becomes
\begin{eqnarray}
p(r) &\propto& e^{-\frac{\lambda}{\beta^2}(r- H/\lambda)^2}.
\end{eqnarray}

\noindent
(B) For $\beta=0$ and $\alpha \neq 0$, Eq. (20) becomes
\begin{eqnarray}
p(r) &\propto& r^{-(2\lambda/ \alpha^2+1)}
e^{-(2H/\alpha^2)/r}.
\end{eqnarray}

Distributions $p(r)$ calculated with the use of Eqs. (22)-(26)
are plotted in Figs. 2(a)-2(c). 
The distribution $p(r)$ for $\alpha=0.0$ 
(without multiplicative noises) in Fig. 2(a)
shows the Gaussian distribution
which is shifted by an applied input $I=0.1$.
When multiplicative noises are added ($\alpha \neq 0$), 
the form of $p(r)$ is changed to the $q$-Gaussian
given by Eq. (22). 
Figure 2(b) shows that when the magnitude of additive noises
$\beta$ is increased,
the width of $p(r)$ is increased.
Figure 2(c) shows that
when the magnitude of external input $I$ is increased, 
$p(r)$ is much shifted and widely spread. 
Note that for $\alpha=0.0$ (no multiplicative noises),
$p(r)$ is simply shifted without a change
in its shape when increasing $I$.

\noindent
{\bf Case II} $F(x)=-\lambda x^a$ and $G(x)=x^b$ ($a, b \ge 0$)

The special case of $a=1$ and $b=1$
has been discussed in the preceding case I [Eqs. (22)-(26)].
For arbitrary $a\ge 0$ and $b \ge 0$,
the probability distribution $p(r)$ given from Eqs. (17)-(19) becomes
\begin{eqnarray}
p(r) \propto 
\left[ 1+ \left( \frac{\alpha^2}{\beta^2}\right) 
\:r^{2b} \right]^{-1/2}\:\exp[X(r)+Y(r)], 
\end{eqnarray}
with
\begin{eqnarray}
X(r)&=& -\left( \frac{2 \lambda r^{a+1}}{\beta^2 (a+1)} \right)
F\left(1,\frac{a+1}{2 b}, \frac{a+1}{2b}+1
;-\frac{\alpha^2r^{2b}}{\beta^{2}} \right), \\
Y(r)&=& \left( \frac{2 H r}{\beta^2} \right)
F\left(1,\frac{1}{2 b}, \frac{1}{2b}+1
;-\frac{\alpha^2r^{2b}}{\beta^{2}} \right), 
\end{eqnarray}
where $F(a,b,c;z)$ is the hypergeometric function.
Some limiting cases of Eqs. (27)-(29) are shown in the following.

\noindent
(a) The case of $H=Y(r)=0$ was previously studied 
in Ref. \cite{Anten02}.

\noindent
(b) For $\alpha=0$ and $\beta \neq0$, we get
\begin{eqnarray}
p(r) &\propto& 
\exp\left[ -\left(\frac{2\lambda}{\beta^2(a+1)} \right)r^{a+1}
+\left( \frac{2 H}{\beta^2} \right) r \right],
\hspace{1cm}\mbox{for $a+1 \neq 0$} \\
&\propto& r^{-2 \lambda/\beta^2} \exp\left(\frac{2H\:r}{\beta^2} \right),
\hspace{4cm}\mbox{for $a+1 = 0$}
\end{eqnarray}

\noindent
(c) For $\beta=0$ and $\alpha \neq0$, we get
\begin{eqnarray}
p(r) &\propto& r^{-b}
\exp\left[ -\left(\frac{2\lambda}{\alpha^2(a-2b+1)} \right) r^{a-2b+1}
-\left( \frac{2H}{\alpha^2(2b-1)} \right) r^{-2b+1} \right], 
\nonumber \\
&&\hspace{6cm}\mbox{for $a-2b+1 \neq 0, 2b-1 \neq 0$} \\
&\propto& r^{-(2\lambda/\alpha^2+b)}
\exp\left[ -\left (\frac{2H}{\alpha^2(2b-1)} \right) r^{-2b+1} \right] ,
\hspace{0.5cm}\mbox{for $a-2b+1 = 0$} \\
&\propto& r^{(2H/\alpha^2-1/2)} 
\exp\left[ -\left( \frac{2\lambda}{\alpha^2 a} \right) r^a \right],
\hspace{1cm}\mbox{for $2b-1 = 0$ ($b=1/2$)} \\
&\propto& r^{-[2(\lambda-H)/\alpha^2+1/2]}, \nonumber \\
&&\hspace{3cm}\mbox{for $a-2b+1 = 0, 2b-1 = 0$ ($a=0$, $b=1/2$)}
\end{eqnarray}

\noindent
(d) In the case of $a=1$ and $b=1/2$, we get
\begin{equation}
p(r) \propto \left( r+\frac{\beta^2}
{\alpha^2} \right)^{(2\lambda \beta^2/\alpha^4+2 H/\alpha^2-1/2)}
\exp\left[ -\left( \frac{2 \lambda}{\alpha^2} \right) r \right],
\end{equation}
which reduces, in the limit of $\alpha=0$, to
\begin{equation}
p(r) \propto \exp\left[-\left( \frac{\lambda}{\beta^2}\right)
\left(r-\frac{H}{\lambda} \right)^2 \right],
\hspace{1cm}\mbox{for $\alpha=0$}
\end{equation}

\noindent
{\bf Case III} $F(x)=-\lambda \ln x$ and $G(x)=x^{1/2}$ ($x > 0$)

We get
\begin{equation}
p(r) \propto r^{-1/2} 
\exp\left[ -\left( \frac{\lambda}{\alpha^2} \right)
\left( \ln r -\frac{H}{\lambda} \right)^2 \right].
\hspace{1cm}\mbox{for $\beta=0$}
\end{equation} 

Figure 3(a) shows distributions $p(r)$ 
for various $a$ with fixed values of 
$I=0.1$, $b=1.0$, $\alpha=1.0$
and $\beta=0.0$ (multiplicative noise only).
With decreasing $a$, a peak of $p(r)$
at $r \sim 0.1$ becomes sharper.
Figure 4(a) shows
distributions $p(r)$
for various $b$ with fixed values of
$I=0.1$, $a=1.0$, $\alpha=1.0$
and $\beta=0.0$ (multiplicative only).
We note that a change in the $b$ value yields 
considerable changes in
shapes of $p(r)$.
Figures 3(b) and 4(b) will be discussed shortly.

\subsubsection{Distribution of $T$}

When the temporal ISI $T$ is simply defined
by $T=1/r$, its distribution $\pi(T)$ 
is given by
\begin{eqnarray}
\pi(T)=p\left( \frac{1}{T} \right) \frac{1}{T^2}.
\end{eqnarray}
%although there is controversy on the relation
%between $r$ and $s$ \cite{Shriki03}.
For $F(x)=-\lambda x$, $G(x)=x$
and $\beta=0$, Eq. (26) or (33) yields
\begin{eqnarray}
\pi(T) \propto
T^{(2\lambda/ \alpha^2-1)}
\exp \left[ -\left(\frac{2H}{\alpha^2} \right) T \right], 
\end{eqnarray}
which expresses the gamma distribution [see Eq. (7)]
\cite{Wilk00,Tuckwell88}.
For $F(x)=-\lambda x^2$, $G(x)=x$ and $\beta=0$, Eq. (32) yields
\begin{eqnarray}
\pi(T) \propto
T^{-1}
\exp\left[-\left( \frac{2H}{\alpha^2} \right)T
- \left( \frac{2\lambda}{\alpha^2} \right) \frac{1}{T} \right], 
\end{eqnarray}
which is similar to the inverse Gaussian distribution
[see Eq. (8)] \cite{Gerstein64}.
For $F(x)=-\ln x$, $G(x)=x^{1/2}$ and $\beta=0$, Eq. (38) leads to
\begin{eqnarray}
\pi(T) \propto
T^{-3/2}
\exp\left[-\left( \frac{2\lambda}{\alpha^2} \right)
\left( \ln T +\frac{H}{\lambda} \right)^2 \right], 
\end{eqnarray}
which is similar to the log-normal distribution [see Eq. (9)]
\cite{McKeegan02}.

Figures 3(b) and 4(b) show distributions of T, $\pi(T)$,
which are obtained from $p(r)$ shown in Figs. 3(a) and
4(a), respectively, by a change of variable [Eq. (39)].
Figure 3(b) shows that with increasing $a$, 
the peak of $\pi(T)$ becomes sharper and moves left. 
We note in Fig. 4(b) that the form of $\pi(T)$ 
significantly varied by changing $b$ in $G(x)=x^b$. 

\subsubsection{Distribution of $R$}

When we consider global variables $R(t)$ defined by
\begin{eqnarray}
R(t)&=&\frac{1}{N} \sum_{i} r_{i}(t),
\end{eqnarray}
the distribution $P(R,t)$ for $R$ is given by
\begin{equation}
P(R,t) = \int \cdots \int \Pi_i \:dr_{i} \:p(\{r_{i} \},t)
\: \delta\Bigl(R-\frac{1}{N}\sum_{j} r_{j}\Bigr).
\end{equation}

Analytic expressions of $P(R)$ are obtainable only for
limited cases.

\noindent
(a) For $\beta \neq 0$ and $ \alpha=0$, $P(R)$ is given by
\begin{eqnarray}
P(R) &\propto& 
\exp\left[ -\left( \frac{\lambda N}{\beta^2} \right)
\left( R- \frac{H}{\lambda} \right)^2 \right], 
\end{eqnarray}
where $H=H(I)$.

\noindent
(b) For $H=0$, we get \cite{Hasegawa06b}
\begin{equation}
P(R)=\frac{1}{2 \pi} \int_{-\infty}^{\infty}\: dk
\;e^{i k R}\:\Phi(k),
\end{equation}
with
\begin{equation}
\Phi(k)=\left[\phi\left( \frac{k}{N} \right) \right]^N,
\end{equation}
where $\phi(k)$ is the characteristic function
for $p(r)$ given by \cite{Abe00}
\begin{eqnarray}
\phi(k)&=& \int_{-\infty}^{\infty} \;
e^{-i k x}\:p(x) dx, \\
&=& 2^{1-\nu}\frac{(\lambda' \mid k \mid )^{\nu}}{\Gamma(\nu)}
K_{\nu}(\lambda' \mid k \mid),
\end{eqnarray}
with
\begin{eqnarray}
\nu&=& \frac{\lambda}{\alpha^2}, \\
\lambda'&=& \frac{\beta}{\alpha}, 
\end{eqnarray}
$K_{\nu}(x)$ expressing the modified Bessel
function.

Some numerical examples of $P(R)$ are
plotted in Figs. 5, 6 and 7 \cite{Note1}.
Figures 5(a) and 5(b) show $P(R)$ for $\alpha=0.0$ 
and $\alpha=0.5$, respectively, when $N$ is changed.
For $\alpha=0.0$, $P(R)$ is the Gaussian whose
width is narrowed by a factor of $1/\sqrt{N}$
with increasing $N$.
In contrast, $P(R)$ for $\alpha=0.5$ is 
the non-Gaussian, whose shape seems to approach
the Gaussian as increasing $N$. These are consistent
with the central-limit theorem.

Effects of an external input $I$ on $p(r)$
and $P(R)$ are examined in Figs. 6(a) and 6(b). 
Figure 6(a) shows that in the case of $\alpha=0.0$
(additive noise only), $p(r)$ and $P(R)$ are simply
shifted by a change in $I$.
This is not the case for $\alpha \neq 0.0$,
for which $p(r)$ and $P(R)$ are shifted and 
widen with increasing $I$, as shown in Fig. 6(b).

Figures 7(a) and 7(b) show effects of the coupling $w$
on $p(r)$ and $P(R)$.
For $\alpha=0.0$, $p(r)$ and $P(R)$ are
little changed with increasing $w$.
On the contrary, for $\alpha = 0.5$,
an introduction of the coupling
significantly modifies $p(r)$ and $P(R)$
as shown in Fig. 7(b).

%\newpage
\subsection{Dynamical property}

\subsubsection{AMM}

Next we will discuss the dynamical property
of the rate model by using the AMM.
Moments of local variables are defined by 
\begin{eqnarray}
\langle r_{i}^k  \rangle &=& 
\int \Pi_{i} \:dr_{i} \: p(\{r_{i} \},t) \:r_{i}^k.
%= \int d\:x_i \:p_i(x_i) \:x_i^k, \\
%\langle R_m^k R_n^{k'} \rangle
%&=& \int \int dR_m \:dR_n \:P(R_m, t) \:P(R_n, t) \:R_m^k \:R_n^{k'}. 
\hspace{1cm}\mbox{($k =1,2,\cdot \cdot$)}
\end{eqnarray}
Equations of motions of means, variances and covariances
of local variables ($r_{i}$) are given by \cite{Hasegawa06a}
\begin{eqnarray}
\frac{d \langle r_{i} \rangle}{dt}
&=& \langle F(r_{i}) \rangle
+\langle H(u_{i}) \rangle
+\frac{\phi \:\alpha^2}{2} \langle G'(r_{i})G(r_{i}) \rangle, \\
\frac{d \langle r_{i} \:r_{j} \rangle}{dt}
&=& \langle r_{i}\:F(r_{j}) \rangle 
+ \langle r_{j}\: F(r_{i}) \rangle 
+ \langle r_{i}\:H(u_{j}) \rangle 
+ \langle r_{j}\: H(u_{i}) \rangle \nonumber \\
&+& \frac{\phi\:\alpha^2}{2} \langle r_{i} G'(r_{j}) G(r_{j}) \rangle
+ \frac{\phi\:\alpha^2}{2} \langle r_{j} G'(r_{i}) G(r_{i})\rangle  
\nonumber \\ 
&+&[\alpha^2\:\langle G(r_{i})^2 \rangle +\beta^2]\:\delta_{ij}. 
\end{eqnarray}
Equations of motions of the mean, variance and covariance
of global variables ($R$) are obtainable 
by using Eqs. (43), (53) and (54): 
\begin{eqnarray}
\frac{d \langle R \rangle}{dt}
&=&\frac{1}{N} \sum_{i} \frac{d \langle r_{i} \rangle}{dt}, \\
\frac{d \langle R^2 \rangle}{dt} 
&=& \frac{1}{N^2}\sum_{i} \sum_{j} 
\frac{d \langle r_{i}\:r_{j} \rangle}{dt}.
\end{eqnarray}

In the AMM \cite{Hasegawa03a},
we define $\mu$, $\gamma$ and $\rho$ given by
\begin{eqnarray}
\mu &=& \langle R \rangle 
= \frac{1}{N} \sum_i \langle r_{i} \rangle, \\
\gamma &=& \frac{1}{N} \sum_i \langle (r_{i}-\mu)^2 \rangle, \\
\rho &=& \langle (R-\mu)^2 \rangle,
\end{eqnarray}
where $\mu$ expresses the mean, $\gamma$ the averaged
fluctuations in local variables ($r_{i}$) 
and $\rho$ fluctuations
in global variable ($R$).
Expanding $r_{i}$ in Eqs. (53)-(56) 
around the average value of $\mu$ as
\begin{equation}
r_{i}=\mu+\delta r_{i},
\end{equation}
and retaining up to the order of 
$<\delta r_{i}\delta r_{j} >$, we get
equations of motions for $\mu$, $\gamma$ and $\rho$ given by 
\begin{eqnarray}
\frac{d \mu}{dt}&=& f_{0}+f_{2}\gamma + h_{0} 
+\left( \frac{\phi \: \alpha^2}{2}\right)
[g_{0}g_{1}+3(g_{1}g_{2}+g_{0}g_{3})\gamma], \\
\frac{d \gamma}{dt} &=& 2f_{1} \gamma
+ 2h_{1} \left( \frac{w N}{Z}\right) 
\left(\rho-\frac{\gamma}{N} \right) \nonumber \\
&+&(\phi+1) (g_{1}^2+2 g_{0}g_{2})\alpha^2\gamma
+ \alpha^2 g_{0}^2+\beta^2, \\
\frac{d \rho}{dt} &=& 2 f_{1} \rho 
+ 2 h_{1} w \rho
+ (\phi+1)
(g_{1}^2+2 g_{0}g_{2})\:\alpha^2 \:\rho 
+ \frac{1}{N}(\alpha^2 g_{0}^2  + \beta^2),
\end{eqnarray}
where 
\begin{eqnarray}
f_{\ell}&=&\frac{1}{\ell !}
\frac{\partial^{\ell} F(\mu)}{\partial x^{\ell}}, \\
g_{\ell}&=&\frac{1}{\ell !}
\frac{\partial^{\ell} G(\mu)}{\partial x^{\ell}}, \\ 
h_{\ell}&=&\frac{1}{\ell !}
\frac{\partial^{\ell} H(u)}{\partial u^{\ell}}, \\
u&=&w \mu +I.
\end{eqnarray}
Original $N$-dimensional stochastic equations given by
Eqs. (10), (11) and (15) are
transformed to the three-dimensional deterministic
equations given by Eqs. (61)-(63).

Before discussing the dynamical property,
we study the stationary property of Eqs. (61)-(63).
In order to make numerical calculations, 
we have adopted 
\begin{eqnarray}
F(x)&=&-\lambda x,  \\
G(x)&=& x, 
\end{eqnarray}
where $\lambda$ stands for the relaxation ratio.
Equations (61)-(63) are expressed 
in the Stratonovich representation by
\begin{eqnarray}
\frac{d \mu}{dt}&=&-\lambda \mu + h_0
%+\frac{c\: h_2 w^2}{Z^2}[N(N-2)\rho+\gamma]
+ \frac{\alpha^2 \mu}{2}, \\
\frac{d \gamma}{dt} &=& -2 \lambda \gamma 
+ \frac{2 h_1 w N}{Z}\left( \rho-\frac{\gamma}{N} \right) 
+ 2 \alpha^2 \gamma + \alpha^2 \mu^2 + \beta^2, \\
\frac{d \rho}{dt} &=& -2\lambda \rho +2 h_1 w \rho
+ 2 \alpha^2 \rho
+ \frac{\alpha^2 \mu^2}{N}  + \frac{\beta^2}{N},
\end{eqnarray}
where $h_0=u/\sqrt{u^2+1}$, 
$h_1=1/(u^2+1)^{3/2}$ and $h_2=-(3\:u/2)/(u^2+1)^{5/2}$.
The stability of the stationary solution given by Eqs. (70)-(72)
may examined by calculating eigenvalues
of their Jacobian matrix, although actual calculations
are tedious.

Figure 8 shows
the $N$ dependences of 
$\gamma$ and $\rho$ in the stationary state
for four sets of parameters:
$(\alpha, \beta, w)=(0.0, 0.1, 0.0)$ (solid curves),
(0.5, 0.1, 0.0) (dashed curves),
(0.0, 0.1, 0.5) (chain curves) and
(0.5, 0.1, 0.5) (double-chain curves),
with $\beta=0.1$, $\lambda=1.0$ and $I=0.1$.
We note that for all the cases, $\rho$ is proportional to
$N^{-1}$, which is easily realized in Eq. (72).
In contrast, $\gamma$ shows a weak $N$ dependence
for a small $N$ ($< 10$).  

\subsubsection{Response to pulse inputs}

We have studied the dynamical property of the rate
model, by applying a pulse input given by
\begin{equation}
I(t)= A \:\Theta(t-t_{1}) \Theta(t_2-t)+I^{(b)},
\end{equation}
with $A=0.5$, $t_1=40$, $t_2=50$ and
$I^{(b)}=0.1$ expressing the background input,
where $\Theta(x)$ denotes the Heaviside function:
$\Theta(x)=1$ for $x \geq 0$ and 0 otherwise.

Figures 9(a), 9(b) and 9(c) show the time dependence of
$\mu$, $\gamma$ and $\rho$
when the input pulse
$I(t)$ given by Eq. (73) is applied:
solid and dashed curves show the results of AMM and 
DS averaged over 1000 trials, respectively,
with $\alpha=0.5$, $\beta=1.0$, $N=10$ and $w=0.5$
\cite{Note1}.
Figures 9(b) and 9(c) show that
an applied input pulse induces changes in
$\gamma$ and $\rho$.
This may be understood from 
$2 \alpha^2$ terms in Eqs. (71) and (72).
The results of AMM shown by solid curves in Figs. 9(a)-(c)
are in good agreement with DS results shown by dashed curves.
Figure 9(d) will be discussed in the followings.

It is possible to discuss the synchrony in a neuronal cluster
with the use of $\gamma$ and $\rho$ 
defined by Eqs. (58) and (59) \cite{Hasegawa03a}.
In order to quantitatively discuss the
synchronization, we first consider the quantity given by
\begin{equation}
P(t)=\frac{1}{N^2} \sum_{i j}<[r_{i}(t)-r_{j}(t)]^2>
=2 [\gamma(t)-\rho(t)].
\end{equation}
When all neurons are in the completely synchronous state,
we get $r_{i}(t)=R(t)$ for all $i$, and 
then $P(t)=0$ in Eq. (74).
On the contrary, we get 
$P(t)=2(1-1/N)\gamma \equiv P_{0}(t)$
in the asynchronous state
where $\rho=\gamma/N$ \cite{Hasegawa03a}.
We may define the synchronization ratio
given by \cite{Hasegawa03a}
\begin{equation}
S(t) =1-\frac{P(t)}{P_{0}(t)}
= \left( \frac{N \rho(t)/\gamma(t)-1}{N-1} \right),
\end{equation}
which is 0 and 1 for completely asynchronous ($P=P_{0}$)  
and synchronous states ($P=0$), respectively.
Figure 9(d) shows
the synchronization ratio $S(t)$ for $\gamma(t)$ and $\rho(t)$
plotted in Figs. 9(b) and 9(c), respectively,
with $\alpha=0.5$, $\beta=1.0$, $N=10$ and $w=0.5$.
The synchronization at $t < 40$ and $t > 60$
%in the absence of an applied input pulse
is 0.15, but it is decreased to 0.03 
at $40 < t < 50$ by an applied pulse.
This is because $\gamma$ is more increased than $\rho$
by an applied pulse.
The synchronization ratio is vanishes for $w=0$,
and it is increased with increasing the coupling
strength \cite{Hasegawa03a}.

Next we show some results when indices $a$ and $b$
in $F(x)=-\lambda x^a$ and $G(x)=x^b$ are changed.
Figure 10(a) shows the time dependence of $\mu$ for
$(a, b)=(1, 1)$ (solid curve) and
$(a, b)=(2, 1)$ (dashed curve) 
with $\alpha=0.0$, $\beta=0.1$, $N=10$ and
$w=0.0$.
The saturated magnitude of $\mu$ for $\alpha=0.5$
is larger than that for $\alpha=0.0$. 
Solid and dashed curves in Fig. 10(b) show $\mu$
for $(a,b)=(1,1)$ and (1,0.5), respectively,
with $\alpha=0.5$, $\beta=0.001$, $N=10$ and $w=0.0$.
Both results show similar responses to an applied
pulse although $\mu$ for a background input of $I^{(b)}=0.1$
for $(a, b)=(1,0.5)$ is a little larger than that
for $(a, b)=(1,1)$.

\subsubsection{Response to sinusoidal inputs}

We have applied also a sinusoidal input given by
\begin{equation}
I(t)= A \: \left[1-\cos \left( \frac{2 \pi t}{T_p} \right)\right]
+I^{(b)},
\end{equation}
with $A=0.5$, $I^{(b)}=0.1$, and $T_p=10$ and 20.
Time dependences of
$\mu$ for $T_p=20$ and $T_p=10$
are plotted in Figs. 11(a) and 11(b), respectively,
with $\alpha=0.5$, $\beta=1.0$, $w=0.0$ and $N=10$,
solid and dashed curves denoting $\mu$ and $I$,
respectively.
The delay time of $\mu$ against an input $I(t)$
is about $\tau_d \sim 1.0$ independent of $T_p$.
The magnitude of $\mu$ for $T_p=10$ is smaller
than that for $T_p=20$.

%\newpage
\section{Multiple neuron clusters}

\subsection{AMM}

We have assumed a neuronal ensemble consisting of
$M$ clusters, whose $m$th cluster includes
$N_m$ neurons.
The dynamics of firing rate $r_{mi}$ ($\geq 0$) of a neuron $i$ 
in the cluster $m$ is assumed to be described
by the Langevin model given by
\begin{eqnarray}
\frac{dr_{mi}}{dt} &=& F(r_{mi}) 
+H(u_{mi})
%+H\left( \frac{u_{mi}-\theta_m}{d_m} \right) 
+ \alpha_m G(r_{mi}) \eta_{mi}(t)+ \beta_m \xi_{mi}(t),\\
&&\hspace{4cm}\mbox{($m=1- M$, $i=1-N_m$)} \nonumber
\end{eqnarray}
with 
\begin{eqnarray}
u_{mi}(t) &=& \left( \frac{w_{mm}}{Z_m} \right) 
\sum_{k (\neq i)} r_{mk}(t)
+\sum_{n(\neq m)} \sum_{\ell}
\left( \frac{w_{mn}}{(M-1) N_n} \right)  r_{n \ell}(t) \nonumber \\
&+& I_m(t), 
\end{eqnarray}
where $F(x)$ and $G(x)$ are arbitrary functions of $x$:
$H(x)$ is given by Eq. (15):
$Z_m$ $(=N_m-1)$ denotes the coordination number: 
$I_m(t)$ expresses an external input to the cluster $m$:
$\alpha_m$ and $\beta_m$ are the strengths of additive and
multiplicative noises, respectively, 
in the cluster $m$ given by 
$\eta_{mi}(t)$ and $\xi_{mi}(t)$ expressing zero-mean Gaussian white
noises with correlations given by
\begin{eqnarray}
<\eta_{ni}(t)\:\eta_{mj}(t')> 
&=& \delta_{nm} \delta_{ij} \delta(t-t'),\\
<\xi_{ni}(t)\:\xi_{mj}(t')> 
&=& \delta_{nm} \delta_{ij} \delta(t-t'),\\
<\eta_{ni}(t)\:\xi_{mj}(t')> &=& 0.
\end{eqnarray}

In the AMM \cite{Hasegawa03a},
we define means, variances and covariances, 
$\mu_m$, $\gamma_m$ and $\rho_{mn}$, given by
\begin{eqnarray}
\mu_m &=& \langle R_m \rangle 
= \frac{1}{N_m} \sum_i \langle r_{mi} \rangle, \\
%
%\gamma &=& \frac{1}{N} \sum_i 
%(\langle x_i^2 \rangle- \langle x_i \rangle^2), \\
\gamma_m &=& \frac{1}{N_m} \sum_i \langle (r_{mi}-\mu_m)^2 \rangle, \\
%
%\rho &=& \langle X^2 \rangle - \langle X \rangle^2.
\rho_{mn} &=& \langle (R_m-\mu_m) (R_n-\mu_n) \rangle,
\end{eqnarray}
where the global variable $R_m$ is defined by 
\begin{eqnarray}
R_m(t)&=&\frac{1}{N_m} \sum_{i} r_{mi}(t).
\end{eqnarray}
%It is noted that $\gamma_m$ expresses the averaged
%fluctuations in local variables ($r_{mi}$) 
%and that $\rho_m$ denotes fluctuations
%in global variable ($R_m$).
Details of deriving equations of motions
for $\mu_m$, $\gamma_m$ and $\rho_{mn}$ are
given in the Appendix [Eqs. (A10)-(A12)].

\subsection{Stationary property}

Now we consider an E-I ensemble ($M=2$)
consisting of excitatory (E) and
inhibitory (I) neuron clusters, for which 
we get equations of motions 
for $\mu_m$, $\gamma_m$ and $\rho_{mn}$
from Eqs. (A10)-(A16):
\begin{eqnarray}
\frac{d \mu_E}{dt}&=& f_{E,0}+f_{E,2}\gamma_E + h_{E,0} \nonumber \\
&+&\left( \frac{\phi \: \alpha_E^2}{2}\right)
[g_{E,0}g_{E,1}+3(g_{E,1}g_{E,2}+g_{E,0}g_{E,3})\gamma_E], \\
\frac{d \mu_I}{dt}&=& f_{I,0}+f_{I,2}\gamma_I + h_{I,0} \nonumber \\
&+&\left( \frac{\phi \: \alpha_I^2}{2}\right)
[g_{I,0}g_{I,1}+3(g_{I,1}g_{I,2}+g_{I,0}g_{I,3})\gamma_I], \\
\frac{d \gamma_E}{dt} &=& 2f_{E,1} \gamma_E
+ 2h_{E,1} \left[ \left( \frac{w_{EE} N_E}{Z_E}\right) 
\left(\rho_{EE}-\frac{\gamma_E}{N_E} \right) 
- w_{EI} \rho_{EI} \right]\nonumber \\
&+&(\phi+1) (g_{E,1}^2+2 g_{E,0}g_{E,2})\alpha_E^2\gamma_E 
+ \alpha_E^2 g_{E,0}^2+\beta_E^2, \\
\frac{d \gamma_I}{dt} &=& 2f_{I,1} \gamma_I
+ 2h_{I,1} \left[ \left( \frac{-w_{II} N_I}{Z_I}\right) 
\left(\rho_{II}-\frac{\gamma_I}{N_I} \right) 
+ w_{IE} \rho_{EI} \right]\nonumber \\
&+&(\phi+1) (g_{I,1}^2+2 g_{I,0}g_{I,2})\alpha_I^2\gamma_I 
+ \alpha_I^2 g_{I,0}^2+\beta_I^2, \\
\frac{d \rho_{EE}}{dt} &=& 2 f_{E,1} \rho_{EE} 
+ 2 h_{E,1} (w_{EE} \rho_{EE}- w_{EI} \rho_{EI}) \nonumber \\
&+&(\phi+1) (g_{E,1}^2+2 g_{E,0}g_{E,2})\:\alpha_E^2 \:\rho_{EE} 
+ \frac{(\alpha_E^2 g_{E,0}^2  + \beta_E^2)}{N_E},\\
\frac{d \rho_{II}}{dt} &=& 2 f_{I,1} \rho_{II} 
+ 2 h_{I,1} (-w_{II} \rho_{II}+ w_{IE} \rho_{EI}) \nonumber \\
&+&(\phi+1) (g_{I,1}^2+2 g_{I,0}g_{I,2})\:\alpha_I^2\: \rho_{II} 
+ \frac{(\alpha_I^2 g_{I,0}^2  + \beta_I^2)}{N_I},\\
\frac{d \rho_{EI}}{dt} &=& (f_{E,1}+f_{I,1}) \rho_{EI} 
\nonumber \\
&+& h_{E,1} (w_{EE} \rho_{EI}-w_{EI} \rho_{II})
+ h_{I,1} (-w_{II} \rho_{EI} + w_{IE} \rho_{EE}) \nonumber \\
&+&\frac{(\phi+1)}{2}
[(g_{E,1}^2+2 g_{E,0}g_{E,2})\:\alpha_E^2 
+ (g_{I,1}^2+2 g_{I,0}g_{I,2})\:\alpha_I^2 ] \rho_{EI}. 
\end{eqnarray}
Here we set $- w_{II} \leq 0$ and $- w_{EI} \leq 0$
after convention.
Equations (86)-(92) correspond to a generalized Wilson-Cowan (WC) 
model, because they reduce to WC model \cite{Wilson72}
if we adopt $F(x)=-\lambda$ and $G(x)=x$ (see below), 
neglecting all fluctuations of $\gamma_{\eta}$ and $\rho_{\eta \eta'}$
($\eta, \eta'$=$E, I$).

It is difficult to obtain the stability condition for 
the stationary solution of Eqs. (86)-(92) because they are 
seven-dimensional nonlinear equations.
We find that equations of motions of
$\mu_E$ and $\mu_I$ are decoupled from the rest of
variables in the cases of $G(x)=x$ and $G(x)=x^{1/2}$ 
with $F(x)=- \lambda \:x$, for which
$f_{\eta,2}=0$ and 
$(g_{\eta,1} g_{\eta,2}+g_{\eta,0} g_{\eta,3})=0$ ($\eta=E,I$)
in Eqs. (86) and (87). 
We will the discuss the stationary solutions
for these two cases in the followings.

\vspace{0.5cm}
\noindent
{\bf (A) $F(x)=-\lambda x$} and {\bf $G(x)=x$}

Equations of motions for $\mu_E$ and $\mu_I$
become
\begin{eqnarray}
\frac{d \mu_E}{dt}
&=&-\left( \lambda_E-\frac{\alpha_E^2}{2} \right) \mu_E
+H(w_{EE} \mu_E-w_{EI} \mu_I+I_E), \\
\frac{d \mu_I}{dt}
&=&- \left( \lambda_I-\frac{\alpha_I^2}{2} \right) \mu_I
+H(w_{IE} \mu_E-w_{II} \mu_I+I_I). 
\end{eqnarray}
The stationary slution is given by
\begin{eqnarray}
0&=& f_1(\mu_E, \mu_I)
=-\left( \lambda_E-\frac{\alpha_E^2}{2} \right) \mu_E
+H(w_{EE} \mu_E-w_{EI} \mu_I+I_E), \\
0&=&f_2(\mu_E, \mu_I)
=-\left( \lambda_I-\frac{\alpha_I^2}{2} \right) \mu_I
+H(w_{IE} \mu_E-w_{II} \mu_I+I_I). 
\end{eqnarray}
The stability condition for stationary solutions is given by
\begin{eqnarray}
T&=&-\lambda_E-\lambda_I+\frac{\alpha_E^2}{2}+\frac{\alpha_I^2}{2}
-w_{EE} h_{E,1} -w_{II} h_{I,1}< 0, \\
D&=& \left( \lambda_E-\frac{\alpha_E^2}{2}-w_{EE}h_{E,1} \right)
\left( \lambda_I-\frac{\alpha_I^2}{2}-w_{II}h_{I,1} \right)
\nonumber \\
&+&w_{EI}w_{IE}h_{E,1}h_{I,1} > 0,
\end{eqnarray}
where $T$ and $D$ denote the trace and determinant, respectively,
of Jacobian matrix of Eqs. (93) and (94).

By solving Eqs. (93) and (94), we get 
stationary solutions of $\mu_E$ and $\mu_I$.
Figures 12(a) and 12(b) show $\mu_E$ and $\mu_I$,
respectively, as a function of $w_{EE}$
for various $\alpha$ ($\equiv \alpha_E=\alpha_I$)
with $\lambda_E=\lambda_I=1.0$,
$w_{EI}=w_{IE}=w_{II}=1$, $I_E=I_I=0$
and $N_E=N_I=10$.
In the case of $\alpha=0$, $\mu_E$ and $\mu_I$
are zero for $w_{EE} \leq w_c$ 
but become finite for $w_{EE} > w_c$ 
where $w_c$ (=1.5) denotes the critical
couplings for the ordered state.
With increasing $\alpha$, 
the critical value of $w_c$ is decreased and
magnitudes of $\mu_E$
and $\mu_I$ in the ordered state are increased. 
This shows that multiplicative noise works to create the
ordered state \cite{Broeck94}.

\vspace{0.5cm}
\noindent
{\bf (B) $F(x)=-\lambda x$} and {\bf $G(x)=x^{1/2}$}

Equations of motions for $\mu_E$ and $\mu_I$
become
\begin{eqnarray}
\frac{d \mu_E}{dt}&=&-\lambda_E \mu_E +\frac{\alpha_E^2}{4}
+H(w_{EE} \mu_E-w_{EI} \mu_I+I_E), \\
\frac{d \mu_I}{dt}&=&-\lambda_I \mu_I +\frac{\alpha_I^2}{4}
+H(w_{IE} \mu_E-w_{II} \mu_I+I_I).
\end{eqnarray}
The stationary solution is given by
\begin{eqnarray}
0&=&f_1(\mu_E, \mu_I)=-\lambda_E \mu_E +\frac{\alpha_E^2}{4}
+H(w_{EE} \mu_E-w_{EI} \mu_I+I_E), \\
0&=&f_2(\mu_E, \mu_I)=-\lambda_I \mu_I +\frac{\alpha_I^2}{4}
+H(w_{IE} \mu_E-w_{II} \mu_I+I_I).
\end{eqnarray}
%\begin{eqnarray}
%\mu_I^{(1)}&=& \frac{1}{w_{EI}}\left(w_{EE} \mu_E+I_E
%-\frac{(\lambda_E \mu_E-\alpha_E^2/4)}
%{\sqrt{1-[(\lambda_E \mu_E-\alpha_E^2/4)/c]^2}} \right), \\
%
%\mu_I^{(2)} &=& \frac{1}{\lambda_I}\left(\frac{w_{IE}\mu_E+I_I}
%{\sqrt{[(w_{IE}\mu_E+I_I)/c]^2+1}}+\frac{\alpha_I^2}{4}\right).
%\end{eqnarray}
The stability condition
for stationary solution is given by
\begin{eqnarray}
T&=&-\lambda_E-\lambda_I
-w_{EE} h_{E,1} -w_{II} h_{I,1}< 0, \\
D&=& (\lambda_E-w_{EE}h_{E,1})
(\lambda_I-w_{II}h_{I,1})+w_{EI}w_{IE}h_{E,1}h_{I,1} > 0,
\end{eqnarray}
where $T$ and $D$ express the trace and determinant, repectively,
of Jacobian matrix of Eqs. (99) and (100).

Figure 13(a) and 13(b) show the $w_{EE}$ dependences
of $\mu_E$ and $\mu_I$, respectively,
for various $\alpha$ ($=\alpha_E=\alpha_I$)
with $\lambda_E=\lambda_I=1.0$,
$w_{EI}=w_{IE}=w_{II}=1.0$ and $N_E=N_I=10$.
Equations (99) and (100) show that multiplicative noise
play a role of inputs, yielding
finite $\mu_E$ and $\mu_I$ even for 
no external inputs ($I_E=I_I=0$).
With increasing $\alpha$, magnitudes of $\mu_E$
and $\mu_I$ are increased and the critical value of $w_c$
for the ordered state is decreased.
It is interesting to note that the behavior of 
$\mu_E$ and $\mu_I$ at $w_{EE} \simg w_c$
is gradually changed with increasing $\alpha$.

\subsection{Dynamical property}

We have studied the dynamical property of the rate
model, by applying pulse inputs given by
\begin{eqnarray}
I_{\eta}(t)= A_{\eta} \:\Theta(t-t_{1}) 
\Theta(t_2-t)+I_{\eta}^{(b)},
\hspace{1cm}\mbox{($\eta=E, I$)}
\end{eqnarray}
with $A_E=0.5$, $A_I=0.3$, $I_E^{(b)}=0.1$, $I_I^{(b)}=0.05$,
$t_1=40$ and $t_2=50$.
The time courses of $\mu_{\eta}$, $\gamma_{\eta}$
and $\rho_{\eta \eta'}$ ($\eta, \eta'=E,I$)
are plotted in Figs. 14(a)-14(c), respectively,
where solid, chain and dashed curves show the results of AMM and
dotted curves those of  
DS averaged over 1000 trials
for $\alpha_E=\alpha_I=0.5$,
$\beta_E=\beta_I=0.1$, 
and $w_{EE}=w_{EI}=w_{IE}=w_{II}=1.0$.
The results of AMM
are in good agreement with DS results.

%In Fig. 14, time courses of $\mu_E$ and $\mu_I$
%for $b=0.5$ of $G(x)=x^b$ are compared to those
%for $b=1.0$ with $F(x)=-\lambda x$
%($w_{EE}=w_{EI}=w_{IE}=w_{II}=1$). 
%Overall responses for the two choices of
%$b$ value are similar, although magnitudes 
%of $\mu_E$ and $\mu_I$ for $b=1/2$ are 
%a little smaller than those for $b=1$.

Responses of $\mu_E$ and $\mu_I$ 
for various sets of couplings
are plotted in Figs. 15(a)-15(f):
$w1001$ in Fig. 15(b), for example, means
that $(w_{EE},w_{EI},w_{IE},w_{II})$ 
=(1,0,0,1).
Figure 15(a) shows the result of no couplings
($w_{EE}=w_{EI}=w_{IE}=w_{II}=0$).
%then their results are the same as those
%of single clusters shown in Fig. 9(a).
When intracluster couplings of
$w_{EE}=1.0$ and $w_{II}=1.0$ are introduced, 
$\mu_I$ in the inhibitory cluster
is much suppressed, while
the excitatory cluster is in the ordered state with
$\mu_E= 0.73$ for no input pulses, 
as shown in Fig. 15(b). 
Figure 15(c) shows that when only 
the intercluster coupling of $w_{EI}$ is introduced,
the magnitude of $\mu_E$ is decreased compared to
that in Fig. 15(a).
In contrast, Fig. 15(d) shows that an addition of only $w_{IE}$
enhances magnitude of $\mu_I$ compared to that in Fig. 15(a).
We note in Fig. 15(e) that when both
intercluster couplings of $w_{EI}$ and
$w_{IE}$ are included, magnitude of $\mu_E$ is considerably
reduced while that of $\mu_I$ is slightly increased.
When all couplings are added, magnitudes of both
$\mu_E$ and $\mu_I$ are increased compared to
those for no couplings shown in Fig. 15(a).
Figure 15(a)-15(f) clearly shows that 
responses of $\mu_E$ and $\mu_I$ to inputs
significantly depend on the couplings.

Figure 16(a) and 16(b) show the synchronization ratios
of $S_E$ and $S_I$, respectively, defined by 
[see Eq. (75)]
\begin{equation}
S_{\eta}(t)
= \left( \frac{N_{\eta} \:\rho_{\eta\eta}(t)
/\gamma_{\eta}(t)-1}{N_{\eta}-1} \right),
\hspace{1cm}\mbox{($\eta=E, I$)}
\end{equation}
for various sets of couplings when inputs
given by Eq. (105) are applied
($A_E=0.5$, $A_I=0.3$, $I_E^{b}=0.1$, $I_I^{b}=0.05$, 
$\alpha_E=\alpha_I=0.5$, $\beta_E=\beta_I=0.1$,
and $N_E=N_I=10$).
First we discuss the synchronization ratio
at the period of $t < 40$ and $t > 60$
where the pulse input is not relevant.
With no intra- and inter-cluster couplings
($w_{EE}=w_{EI}=w_{IE}=w_{II}=0$), we get $S_E=S_I=0$. 
When only the intra-cluster couplings
of $w_{EE}=1$ and $w_{II}=1$ are introduced, we get
$S_E=0.15$ and $S_I=-0.67$, as shown by dashed curve in Fig. 16(b).
When inter-cluster coupling of $w_{EI}=1$ is included,
the synchronization in the excitatory cluster is decreased to
$S_E=0.08$ (dotted curves in Fig. 16(a). In contrast,
when inter-cluster of $w_{IE}=1$ is introduced, 
the synchrony in the inhibitory cluster is
increased to $S_I=0.06$, as shown by
dotted curve in Fig. 16(b).
When inter-cluster couplings of $w_{EI}=w_{IE}=1$
are included, the synchronization ratios almost vanish
(chain curves).
Solid curves show that
when both intra- and inter-cluster couplings are included,
we get $S_E=0.24$ and $S_I=0.04$.
It is noted that
the responses of the synchronizations to a pulse
applied at $40 \leq t < 50$
are rather complicated. When an input pulse is 
applied at $t=40$, the synchronization ratios are generally
decreased while $S_E$ with $w0110$ increased:
$S_E$ with $w0100$ is once decreased and then increased.
When an applied pulse disappears at $t = 50$,
the synchronization are increased 
in the refractory period
though the synchronization ratios 
for $w0100$ and $w0110$ are decreased.

Figures 17(a)-17(e) show responses of an E-I ensemble
with $N_E=N_I=10$ when the input pulse 
shown in 17(a) is applied only to the excitatory cluster.
Responses of the local rate of $r_{\eta}$
of single neurons in the excitatory ($\eta=E$) 
and inhibitory clusters ($\eta=I$)
are shown by solid and dashed curves, respectively,
in Fig. 17(b).
Local rates in Fig. 17(b)
which are obtained by DS with a single trial,
have much irregularity
induced by additive and multiplicative noises
with $\alpha_E=\alpha_I=0.1$ and $\beta_E=\beta_I=0.1$.
Figure 17(c) shows the population-averaged rates of
$R_{\eta}(t)$ obtained by DS with a single trial.
The irregularity shown in Fig. 17(c) is reduced
compared to that of $r_{\eta}(t)$ in Fig. 17(b),
which demonstrates the advantage of 
the population code.
When DS is repeated and the global variable is averaged 
over 100 trials, we get the result of $R_{\eta}(t)$
whose irregularity is
much reduced by the average over trials,
as shown in Fig. 17(d).
The AMM results of $\mu_{\eta}(t)$ shown in Fig. 17(e)
are in good agreement with those shown in Fig. 17(d).

%\newpage
\section{Discussion and conclusion}

We may calculate the stationary distributions 
in the E-I ensemble.
Figures 18(a)-18(c) show global distributions of
$P_E(R)$ and $P_I(R)$ which are averaged
within the excitatory and
inhibitory clusters, respectively.
We have studied how distributions are varied
when couplings are changed: $w0110$ 
in Fig. 18(b), for example, means
$(w_{EE}, w_{EI}, w_{IE}, w_{II})=(0,1,1,0)$.
Figure 18(a) shows the results without couplings,
for which distributions of $P_E(R)$ and $P_I(R)$
have peaks at $R \sim 0.1$ and 0.05, respectively,
because of applied background inputs 
($I_E^{(b)}=0.1$ and $I_I^{(b)}=0.05$).
When intercluster couplings of $w_{EI}=1.0$ and $w_{IE}=1.0$
are introduced, the peak of $P_E(R)$ locates 
at $R \sim 0.02$, while that of $P_I(R)$ is
at $R \sim 0.08$: their relative positions
are interchanged compared to the case of no couplings
shown in Fig. 18(a). 
When all couplings with $w_{EE}=w_{EI}=w_{IE}=w_{II}=1.0$
are included, we get the distributions shown in Fig. 18(c),
where both $P_E(R)$ and $P_I(R)$ 
have wider distributions than those
with no couplings shown in Fig. 18(a).
Our calculations show that the distributions are much 
influenced by the magnitudes of couplings.

We have proposed the rate model given by Eqs. (10) and (11),
in which the relaxation process is given by
a single $F(x)$.
Instead, when the relaxation process consists of two terms:
\begin{equation}
F(x) \rightarrow c_1 F_1(x)+c_2 F_2(x),
\end{equation}
with $c_1+c_2=1$,
the distribution becomes
\begin{equation}
p(r)=[p_1(r)]^{c_1}\:[p_2(r)]^{c_2},
\end{equation}
where $p_k(r)$ ($k=1, 2$) denotes the distribution only
with $F(x)=F_1(x)$ or $F(x)=F_1(x)$.
In contrast, when multiplicative noises
arise from two independent origins:
\begin{equation}
\alpha x \eta(t) 
\rightarrow c_1 \alpha_1 x \eta_1(t)
+c_2 \alpha_2 x \eta_2(t),
\end{equation}
the distribution for $\beta=H=0$ becomes 
\begin{equation}
p(r) \propto r^{-[2 \lambda/(c_1 \alpha_1^2+c_2 \alpha_2^2)+1]}.
\end{equation}
Similarly, when additive noises
arise from two independent origins:
\begin{equation}
\beta \xi(t) 
\rightarrow c_1 \beta_1 \xi_1(t)+c_2 \beta_2 \xi_2(t),
\end{equation}
the distribution for $\alpha=H=0$ becomes 
\begin{equation}
p(r) \propto {\rm e}^{-\lambda/(c_1 \alpha_1^2+c_2 \alpha_2^2)}.
\end{equation}
Equations (108), (110) and (112) are quite different from the
form given by
\begin{equation}
p(r)=c_1 p_1(r)+c_2 p_2(r),
\end{equation}
which has been conventionally adopted for a fitting of
theoretical distributions to that obtained by experiments.

It is an interesting subject to decode the stimulus
from the observed spiking rate of neurons.
In BMI, stimulus ${\bf \hat{y}}(t)$ 
is decoded from observed rate signals 
${\bf \hat{r}}(t-u)$ with the use of Eq. (4).
In the AMM, the relation 
between the input $I(t)$ and the averaged rate of $R(t)$,
$\mu(t)$, given by Eq. (70) yields that 
$I(t)$ is expressed in terms of $\mu(t)$ and $d \mu(t)/d t$
as 
\begin{eqnarray}
I(t)  &=& \frac{[d\mu/dt+(\lambda-\alpha^2/2)\mu(t)]}
{\sqrt{1-[d\mu/dt+(\lambda-\alpha^2/2)\mu(t)]^2}} - w \mu(t), \\
&\sim& \frac{d\mu}{dt}
+\left( \lambda-\frac{\alpha^2}{2}-w \right) \mu(t).
\hspace{1cm}\mbox{for small $\mu(t)$}
\end{eqnarray}
%where $\langle \cdot \rangle_I$ stands for the average over input $I$.
The $d \mu/dt$ term plays an important role in decoding
dynamics of $\mu(t)$.
%Similarly, $(\lambda-\alpha^2/2-w ) \sqrt{ \rho(t)}$
%provides us with the width of $P(I \mid R)$.
In an approach using the Bayesian statistics \cite{Bayesian}, 
the conditional probability of an input $I$
for a given rate $R$, $P(I \mid R)$, is expressed by 
\begin{eqnarray}
P(I \mid R) \propto P(R \mid I)\: P(I),
\end{eqnarray}
where $P(R \mid I)$ is the conditional
probability of $R$ for a given $I$
and $P(I)$ the likelihood of $I$. 
An estimation of the value of $I$ which yields 
the maximum of $P(I \mid R)$,
is known as the maximum a posteriori (MAP) estimate. 
Since $P(R \mid I)$ is obtainable with the use
of our rate model, as shown in Figs. 6(a) and 6(b),
we may estimate $P(I \mid R)$ by Eq. (116) 
if $P(I)$ is provided.
We note that $I(t)$ given by Eq. (114) corresponds to
the center of gravity of $P(I \mid R)$,
which is expected to be
nearly the same as the maximum value obtained by MAP estimate. 
More sophisticate Bayesian approach using the recursive
method has been proposed \cite{Bayesian}, 
although it is not unclear whether such Bayesian networks
may be implemented by real neurons. 

The structures of neuronal networks 
have been discussed
based on the theory on complex networks
\cite{Watts98}\cite{Bara99}.
The neural network of nematode worm {\it C. elegans}
is reported to be small-world network.
It has been recently observed by 
the functional magnetic resonance imaging (fMRI)
that the functional connectivity in human brain
has the behavior of the scale-free \cite{Chialvo04}
or small-world network \cite{Stam04}\cite{Achard06}.
Most of theoretical studies have assumed the local
or all-to-all coupling in neuron networks,
as we have made in our study.
In real neural networks, however, the couplings are
neither local nor global. 
A new approach extending the AMM 
has been proposed to take into account
the couplings from local to global and/or from
regular to random couplings
\cite{Hasegawa04c}.
It has been shown that the synchronization in the 
small-world networks is worse than that in the
regular networks due to the randomness
introduced in the small-world networks.
%against our common wisdom. 

In recent years, it becomes increasingly popular
to study the distributed information processing
by using cultured neuronal networks which are
cultivated in an artificial way \cite{Culnet}.
It is possible that every cell in the cultured 
network may be observed, monitored, stimulated
and manipulated with high temporal 
and spatial resolutions.
The observed ISI distributions of the cultured networks
with 50-$10^6$ neurons are reported to obey the 
scale-free distribution \cite{Volman04}.
Although our AMM study in Sec. 3 has been made
for an E-I ($M=2$) cluster, 
it would be interesting to investigate the 
dynamics of larger ensembles modeling complex networks
and cultured networks, % by using the new method \cite{Hasegawa04c},
which is left for our future study.

To summarize,
we have discussed the stationary and dynamical properties
of the generalized rate model
by using the FPE and AMM.
The proposed rate model is a phenomenological one and
has no biological basis.
Nevertheless, the generalized
rate model is expected to be useful
in discussing various
properties of neuronal ensembles.
Indeed, the proposed rate model has an interesting property,
yielding various types of
stationary non-Gaussian distributions
such as gamma, inverse-Gaussian 
and log-normal distributions,
which have been experimentally observed
\cite{Kass05}-\cite{McKeegan02}.
The stationary distribution and dynamical responses
of neuronal clusters have been shown to
considerably depend on the model parameters
such as strengths of noises and couplings.
A disadvantage of our AMM is that its applicability 
is limited to weak-noise cases. 
On the contrary, an advantage of the AMM is that we can easily discuss
dynamical property of an $N$-unit neuronal cluster.
In DS and FPE, we have to solve
the $N$-dimensional stochastic Langevin equations and
the $(N+1)$-dimensional partial differential 
equations, respectively, which are more laborious than
the three-dimensional ordinary differential equations
in the AMM.
We hope that the proposed rate model in the AMM is adopted
for a wide class of study on neuronal ensembles.

%The AMM may be applied not only to neural models but also
%to wide classes of stochastic ensembles.

\section*{Acknowledgements}
This work is partly supported by
a Grant-in-Aid for Scientific Research from the Japanese 
Ministry of Education, Culture, Sports, Science and Technology.  

%\newpage
\vspace{1cm}

\noindent
{\Large\bf APPENDIX}
\appendix\section{AMM for multiple clusters}

We will present a detail of an application of
the AMM to the rate model
describing multiple clusters given by Eqs. (77) and (78).
The Fokker-Planck equation for the distribution of
$p(\{r_{mi} \},t)$ %($= p$) 
is given by 
\cite{Haken83}
\begin{eqnarray}
&&\frac{\partial}{\partial t}\: p(\{r_{mi} \},t) \nonumber \\
&&=
-\sum_{mk} \frac{\partial}{\partial r_{mk}}\{ [F(r_{mk}) +
\frac{\phi\alpha_m^2}{2}G'(r_{mk})G(r_{mk}) 
+ H(u_{mk})]\:p(\{ r_{mi} \},t)\}  
\nonumber \\
&&+\frac{1}{2}\sum_{mk}\frac{\partial^2}{\partial r_{mk}^2} 
\{[\alpha_m^2 G(r_{mk})^2+\beta_m^2]\:p(\{ r_{mi} \},t) \},
\nonumber \hspace{3cm}\mbox{(A1)}
\end{eqnarray}
where $G'(x)=dG(x)/dx$, and
$\phi=1$ and 0 in the Stratonovich and Ito representations,
respectively.

When we consider global variables of the
cluster $m$ given by
\begin{eqnarray}
R_m(t)&=&\frac{1}{N_m} \sum_{i} r_{mi}(t),
\nonumber \hspace{7cm}\mbox{(A2)}
\end{eqnarray}
the distribution $P(R_m,t)$ for $R_m$ is given by
\begin{eqnarray}
P(R_m,t) &=& \int \cdots \int \Pi_i \:dr_{mi} \:p(\{r_{mi} \},t)
\: \delta\Bigl(R_m-\frac{1}{N_m}\sum_{j} r_{mj}\Bigr).
\nonumber \hspace{1cm}\mbox{(A3)}
\end{eqnarray}

Variances and covariances of local variables are defined by 
\begin{eqnarray}
\langle r_{mi}^k r_{nj}^{k'} \rangle &=& 
\int \Pi_{mi} \:dr_{mi} \: p(\{r_{mi} \},t) \:r_{mi}^k \:r_{nj}^{k'}. 
%= \int d\:x_i \:p_i(x_i) \:x_i^k, \\
%\langle R_m^k R_n^{k'} \rangle
%&=& \int \int dR_m \:dR_n \:P(R_m, t) \:P(R_n, t) \:R_m^k \:R_n^{k'}. 
\hspace{1cm}\mbox{($k,\:k'=1,2,\cdot \cdot$)}
\nonumber \hspace{0.5cm}\mbox{(A4)}
\end{eqnarray}
Equations of motions of means, variances and covariances
of local variables ($r_{mi}$) are given by
\begin{eqnarray}
\frac{d \langle r_{mi} \rangle}{dt}
&=& \langle F(r_{mi}) \rangle
+\langle H(u_{mi}) \rangle
+\frac{\phi \:\alpha_m^2}{2} \langle G'(r_{mi})G(r_{mi}) \rangle, 
\nonumber \hspace{2cm}\mbox{(A5)}\\
\frac{d \langle r_{mi} \:r_{nj} \rangle}{dt}
&=& \langle r_{mi}\:F(r_{nj}) \rangle 
+ \langle r_{nj}\: F(r_{ni}) \rangle 
+ \langle r_{mi}\:H(u_{nj}) \rangle 
+ \langle r_{nj}\: H(u_{mi}) \rangle \nonumber \\
&+& \frac{\phi\:\alpha_n^2}{2} \langle r_{mi} G'(r_{nj}) G(r_{nj}) \rangle
+ \frac{\phi\:\alpha_m^2}{2} \langle r_{nj} G'(r_{mi}) G(r_{mi})\rangle  \nonumber \\ 
&+&[\alpha_m^2\:\langle G(r_{mi})^2 \rangle +\beta_m^2]
\:\delta_{ij}\delta_{mn}. 
\nonumber \hspace{5cm}\mbox{(A6)}
\end{eqnarray}
Equations of motions of the mean, variance and covariance
of global variables ($R_m$) are obtainable 
by using Eqs. (A3), (A5) and (A6): 
\begin{eqnarray}
\frac{d \langle R_m \rangle}{dt}
&=&\frac{1}{N_m} \sum_{i} \frac{d \langle r_{mi} \rangle}{dt}, 
\nonumber \hspace{5cm}\mbox{(A7)} \\
\frac{d \langle R_m R_n \rangle}{dt} 
&=& \frac{1}{N_m N_n}\sum_{i} \sum_{j} 
\frac{d \langle r_{mi}\:r_{nj} \rangle}{dt}.
\nonumber \hspace{3cm}\mbox{(A8)}
\end{eqnarray}

Variances and covariances, 
$\mu_m$, $\gamma_m$ and $\rho_{mn}$ are given by
Eqs. (82)-(84).
Expanding $r_{mi}$ in Eqs. (A5)-(A8) 
around the average value of $\mu_m$ as
\begin{eqnarray}
r_{mi}&=&\mu_m+\delta r_{mi},
\nonumber \hspace{5cm}\mbox{(A9)}
\end{eqnarray}
and retaining up to the order of 
$<\delta r_{mi}\delta r_{mj} >$, we get
equations of motions for 
$\mu_m$, $\gamma_m$ and $\rho_{mn}$ given by 
\begin{eqnarray}
\frac{d \mu_m}{dt}&=& f_{m,0}+f_{m,2}\gamma_m + h_{m,0} \nonumber \\
&+&\left( \frac{\phi \: \alpha_m^2}{2}\right)
[g_{m,0}g_{m,1}+3(g_{m,1}g_{m,2}+g_{m,0}g_{m,3})\gamma_m], 
\nonumber \hspace{2cm}\mbox{(A10)}
\\
\frac{d \gamma_m}{dt} &=& 2f_{m,1} \gamma_m
+ 2h_{m,1} \left[ \left( \frac{w_{mm} N_m}{Z_m}\right) 
\left(\rho_{mm}-\frac{\gamma_m}{N_m} \right) 
+ \left( \frac{1}{M-1} \right)
\sum_{n (\neq m)} w_{mn} \rho_{mn} \right]\nonumber \\
&+&(\phi+1) (g_{m,1}^2+2 g_{m,0}g_{m,2})\alpha_m^2\gamma_m 
+ \alpha_m^2 g_{m,0}^2+\beta_m^2, 
\nonumber \hspace{2cm}\mbox{(A11)} \\
\frac{d \rho_{mn}}{dt} &=& (f_{m,1}+f_{n,1}) \rho_{mn} 
+ h_{m,1} \left[w_{mm} \rho_{mn}
+\left( \frac{1}{M-1} \right)
\sum_{n' (\neq m)} w_{mn'} \rho_{nn'} \right] \nonumber \\
&+& h_{n,1} \left[w_{nn} \rho_{mn}
+ \left( \frac{1}{M-1} \right)
\sum_{n' (\neq n)} w_{nn'} \rho_{mn'} \right] \nonumber \\
&+&\frac{(\phi+1)}{2}
[(g_{m,1}^2+2 g_{m,0}g_{m,2})\:\alpha_m^2 
+ (g_{n,1}^2+2 g_{n,0}g_{n,2})\:\alpha_n^2 ] \rho_{mn} 
\nonumber \\
&+& \delta_{mn}\left(\frac{\alpha_m^2 g_{m,0}^2  + \beta_m^2}{N_m}\right),
\nonumber \hspace{6cm}\mbox{(A12)}
\end{eqnarray}
where 
\begin{eqnarray}
f_{m,\ell}&=&\frac{1}{\ell !}
\frac{\partial^{\ell} F(\mu_m)}{\partial x^{\ell}}, \nonumber \\
g_{m,\ell}&=&\frac{1}{\ell !}
\frac{\partial^{\ell} G(\mu_m)}{\partial x^{\ell}}, \nonumber\\ 
h_{m,\ell}&=&\frac{1}{\ell !}
\frac{\partial^{\ell} H(u_m)}{\partial u^{\ell}},\nonumber\\
u_m&=&w_{mm}\mu_m
+\left( \frac{1}{M-1} \right) 
\sum_{n (\neq m)} \:w_{mn} \mu_n+I_m.\nonumber
\end{eqnarray}
For a two-cluster ($M=2$) ensemble consisting 
of excitatory and inhibitory clusters, 
equations of motions given by Eqs. (A10)-(A12)
reduce to Eqs. (86)-(92).

\newpage
%\begin{references}

%\end{references}

\newpage

\begin{figure}
\caption{
%Fig W.
Response of a 10-unit FN neuron cluster 
subjected to additive and multiplicative noises;
(a) input signal $I(t)$,
(b) a local membrane potential [$v(t)$] and
(c) a global membrane potential [$V(t)=(1/N)\sum_i v_i(t)$] 
obtained by direct simulation (DS) with a single trial:
(d) a global membrane potential [$V(t)$] 
calculated by DS with 100 trials:
(e) the result of the AMM [$\mu(t)$].
%(see the Appendix:
%$\alpha=0.1$, $\beta=0.2$, $J=0.5$).
}
\label{fig1}
\end{figure}

\begin{figure}
\caption{
%Fig A.
(a) Distributions $p(r)$ of the (local) firing rate $r$
for various $\alpha$ with $\lambda=1.0$, 
$\beta=0.1$, $I=0.1$ and $w=0.0$,
(b) $p(r)$
for various $\beta$ with $\lambda=1.0$, 
$\alpha=1.0$, $I=0.1$ and $w=0.0$, and
(c) $p(r)$
for various $I$ with $\lambda=1.0$, 
$\alpha=0.5$, $\beta=0.1$ and $w=0.0$.
}
\label{fig2}
\end{figure}

\begin{figure}
\caption{
%Fig F.
(a) Distributions $p(r)$ of the (local) firing rate $r$
and (b) $\pi(T)$ of the ISI $T$
for $a=0.8$ (chain curves), $a=1.0$ (solid curves),
$a=1.5$ (dotted curves) and $a=2.0$ (dashed curves)
with $b=1.0$, $\lambda=1.0$, 
$\alpha=1.0$, $\beta=0.0$ and $I=0.1$.
}
\label{fig3}
\end{figure}

\begin{figure}
\caption{
%Fig G.
(a) Distributions $p(r)$ of the (local) firing rate $r$
and (b) $\pi(T)$ of the ISI $T$
for $b=0.5$ (dashed curves),
$b=1.0$ (solid curves), $b=1.5$ (dotted curves)
and $b=2.0$ (chain curves) 
with $a=1.0$, $\lambda=1.0$, 
$\alpha=1.0$, $\beta=0.0$ and $I=0.1$:
results for $b=1.5$ and $b=2$ should be multiplied 
by factors of 2 and 5, respectively.
}
\label{fig4}
\end{figure}

\begin{figure}
\caption{
%Fig E.
Distributions $P(R)$ of the (global) firing rate $R$
for (a) $\alpha=0.0$
and (b) $\alpha=0.5$, with $N=1$, 10 and 100:
$\lambda=1.0$, $\beta=0.1$, $w=0.0$
and $I=0.1$.
}
\label{fig5}
\end{figure}

\begin{figure}
\caption{
%Fig C.
Distributions $p(r)$ (dashed curves)
and $P(R)$ (solid curves)
for (a) $\alpha=0.0$ and (b) $\alpha=0.5$ 
with $I=0.1$ and $I=0.2$:
$N=10$, $\lambda=1.0$, $\beta=0.1$ and $w=0.0$.
}
\label{fig6}
\end{figure}

\begin{figure}
\caption{
%Fig D.
Distributions $p(r)$ (dashed curves)
and $P(R)$ (solid curves)
for (a) $\alpha=0.0$ and (b) $\alpha=0.5$
with $w=0.0$ and $w=0.5$:
$N=10$, $\lambda=1.0$, $\beta=0.1$ and $I=0.1$.
}
\label{fig7}
\end{figure}

\begin{figure}
\caption{
%Fig I.
The $N$ dependence of 
$\gamma$ and $\rho$ in the stationary states 
for four sets of parameters:
$(\alpha, \beta, w)=(0.0, 0.1, 0.0)$ (solid curves),
$(0.5, 0.1, 0.0)$ (dashed curves),
$(0.0, 0.1, 0.5)$ (chain curves) and
$(0.5, 0.1, 0.5)$ (double-chain curves):
$\lambda=1.0$, $N=10$ and $I=0.1$.
}
\label{fig8}
\end{figure}

\begin{figure}
\caption{
%Fig J.
Time courses of 
(a) $\mu(t)$, (b) $\gamma(t)$, (c) $\rho(t)$
and (d) $S(t)$
for a pulse input $I(t)$ given by Eq. (73)
with $\lambda=1.0$,
$\alpha=0.5$, $\beta=0.1$, $N=10$ and $w=0.5$,
solid and chain curves 
denoting results of AMM and dashed curves expressing
those of DS result with 1000 trials.
}
\label{fig9}
\end{figure}

\begin{figure}
\caption{
%Fig H.
(a) Response of $\mu(t)$ to input pulse $I(t)$ 
given by Eq. (73)
for $(a,b)=(1, 1)$ (solid curve) and 
$(a, b)=(2, 1)$ (dashed curve) 
with $\alpha=0.0$, $\beta=0.1$,
$N=10$ and $\lambda=1.0$.
(b) Response of $\mu(t)$ to input pulse $I(t)$
for $(a,b)=(1, 1)$ (solid curve) and 
$(a, b)=(1, 0.5)$ (dashed curve) 
with $\alpha=0.5$, $\beta=0.001$,
$N=10$, $\lambda=1.0$ and $w=0.0$.
}
\label{fig10}
\end{figure}

\begin{figure}
\caption{
%Fig K.
Response of $\mu(t)$ (solid curves)
to sinusoidal input $I(t)$ (dashed curves)
given by Eq. (73)
for (a) $T_p=20$ and (b) $T_p=10$
with $A=0.5$, $\lambda=1.0$, $\alpha=0.5$, 
$\beta=0.1$, $w=0$ and $N=10$ 
($a=1$ and $b=1$).
}
\label{fig11}
\end{figure}

\begin{figure}
\caption{
%Fig M.
Stationary values of (a) $\mu_E$ and (b) $\mu_I$
in an E-I ensemble
as a function of $w_{EE}$ for various values
of $\alpha$ ($=\alpha_E=\alpha_I$)
for the case of $G(x)=x$
with $\lambda_E=\lambda_I=1.0$,
$w_{EI}=w_{IE}=w_{II}=1.0$ and $I_E=I_I=0$
and $N_E=N_I=10$.
}
\label{fig12}
\end{figure}

\begin{figure}
\caption{
%Fig N.
Stationary values of (a) $\mu_E$ and (b) $\mu_I$
in an E-I ensemble
as a function of $w_{EE}$ for various values
of $\alpha$ ($=\alpha_E=\alpha_I$)
for the case of $G(x)=x^{1/2}$
with $\lambda_E=\lambda_I=1.0$,
$w_{EI}=w_{IE}=w_{II}=1.0$ and $I_E=I_I=0$
and $N_E=N_I=10$.
}
\label{fig13}
\end{figure}

\begin{figure}
\caption{
%Fig J.
Time courses of 
(a) $\mu_E$ and $\mu_I$, 
(b) $\gamma_E$ and $\gamma_I$, 
and (c) $\rho_{EE}$, $\rho_{II}$ and $\rho_{EI}$,
for pulse inputs given by Eq. (105)
with $\alpha_E=\alpha_I \:(= \alpha)=0.5$: 
$\beta_E=\beta_I \:(= \beta)=0.1$, $N_E=N_I=10$ and 
$w_{EE}=w_{EI}=w_{IE}=w_{II}=1.0$:
solid, dashed and chain curves denote
the results of AMM and dotted curves
express those of DS with 1000 trials. 
}
\label{fig14}
\end{figure}

\begin{figure}
\caption{
%Fig P.
Responses of $\mu_E$ (solid curves) and 
$\mu_I$ (dashed curves) of an E-I ensemble
to pulse inputs given by Eq. (105);
(a) $(w_{EE}, w_{EI}, w_{IE}, w_{II})$=(0,0,0,0),
(b) (1,0,0,1),
(c) (0,1,0,0),
(d) (0,0,1,0),
(e) (0,1,1,0)  and
(f) (1,1,1,1):
$w1001$, for example, expresses the case of (b):
$\alpha_E=\alpha_I=0.5$, $\beta_E=\beta_I=0.1$,
$A_E=0.5$, $A_I=0.3$, $I_E^{(b)}=0.1$, $I_I^{(b)}=0.05$ 
and $N_E=N_I=10$.
}
\label{fig15}
\end{figure}

\begin{figure}
\caption{
%Fig P.
Synchronization ratios of (a) $S_E$ and 
(b) $S_I$ of an E-I ensemble
to pulse inputs given by Eq. (105)
with $A_E=0.5$, $A_I=0.3$, $I_E^{(b)}=0.1$, $I_I^{(b)}=0.05$, 
$\alpha_E=\alpha_I=0.5$, $\beta_E=\beta_I=0.1$,
and $N_E=N_I=10$:
$w1001$, for example, denotes
$(w_{EE}, w_{EI}, w_{IE}, w_{II})$=(1,0,0,1).
}
\label{fig16}
\end{figure}

\begin{figure}
\caption{
%Fig Y.
Responses of an E-I ensemble;
(a) input signal $I_E(t)$,
(b) local rates $r_{\eta}(t)$ and
(c) global rate $R_{\eta}(t)$ ($\eta=E$ and $I$) 
obtained by direct simulation (DS) with a single trial:
(d) global rates $R_{\eta}(t)$
calculated by DS with 100 trials:
(e) the result of the AMM $\mu_{\eta}(t)$:
$\alpha_E=\alpha_I=0.5$, $\beta_E=\beta_I=0.1$,
$w_{EE}=w_{EI}=w_{IE}=w_{II}=1.0$,
$A_E=0.5$, $A_I=0.0$, $I_E^{(b)}=0.1$, $I_I^{(b)}=0.05$ 
and $N_E=N_I=10$.
Solid and dashed curves in (b)-(e) denote
the results for excitatory (E) and inhibitory (I) clusters,
respectively.
%(see the Appendix:
%$\alpha=0.1$, $\beta=0.2$, $J=0.5$).
}
\label{fig17}
\end{figure}

\begin{figure}
\caption{
%Fig U.
Stationary global distributions 
in E [$P_E(R)$] and I clusters [$P_I(R)$] 
for (a) $(w_{EE}, w_{EI}, w_{IE}, w_{II})$=(0,0,0,0),
(b) (0,1,1,0) and (c) (1,1,1,1)
with $I_E^{(b)}=0.1$, $I_I^{(b)}=0.05$,
$\alpha_E=\alpha_I\:(=\alpha)=0.5$, 
$\beta_E=\beta_I\:(=\beta)=0.1$
and $N_E=N_I=10$,
solid and dashed curves denoting $P_E(R)$
and $P_I(R)$, respectively.
}
\label{fig18}
\end{figure}

\end{document}